\def\etal{{et al. }}

\def\tx{T_X}
\def\mvir{M_{\rm vir}}
\def\mvirtx{M_{\rm vir} {\rm -} T_X}
\def\rhocr{\rho_{\rm cr}}

\def\Omat{\Omega_{\rm M}}
\def\Olam{\Omega_{\Lambda}}

\documentstyle[12pt,aaspp4,epsf]{article}
\begin{document}
\lefthead{VOIT}
\righthead{CLUSTER TEMPERATURE EVOLUTION}
\slugcomment{To appear in \apj , 20 Nov 2000}
\title{Cluster Temperature Evolution: The Mass-Temperature Relation}
\author{G. Mark Voit\footnote{voit@stsci.edu} } 
\affil{Space Telescope Science Institute \\3700 San Martin Drive \\
Baltimore, MD 21218 }

\begin{abstract}
Evolution of the cluster temperature function is extremely
sensitive to the mean matter density of the universe.
Current measurements based on cluster temperature surveys
indicate that $\Omat \approx 0.3$ with a $1\sigma$ statistical
error $\sim 0.1$, but the systematic errors in this method
are of comparable size.  Many more high-$z$ cluster temperatures 
will be arriving from {\em Chandra} and {\em XMM} in the near
future.  In preparation for future cluster temperature surveys,
this paper analyses the cluster mass-temperature relation, with
the intention of identifying and reducing the systematic errors
it introduces into measurements of cosmological parameters.
We show that the usual derivation of this relation from spherical
top-hat collapse is physically inconsistent and propose a
more realistic derivation based on a hierarchical merging model
that more faithfully reflects the gradual ceasing of cluster evolution
in a low-$\Omat$ universe.
We also analyze the effects of current systematic uncertainties 
in the $\mvirtx$ relation and show that they introduce a systematic
uncertainty of $\sim 0.1$ in the best-fitting $\Omat$.  Future
improvements in the accuracy of the $\mvirtx$ relation will
most likely come from comparisons of predicted cluster temperature
functions with temperature functions derived directly from 
large-scale structure simulations.
\end{abstract}


\section{Introduction}

Surveys of distant clusters of galaxies are now realizing their
promise as cosmological indicators (e.g., Henry 1997; Eke \etal 1998;
Borgani \etal 1999a; Donahue \& Voit 1999).  Because clusters are
the largest virialized objects in the universe, and the latest
objects to form in hierarchical models of structure formation,
their rate of evolution is quite sensitive to cosmological 
parameters.  However, because cluster masses are difficult 
to measure directly, a surrogate for cluster mass is usually
used when comparing cluster observations to structure-formation
models. In the X-ray regime, the simplest cluster observables 
are X-ray luminosity ($L_X$) and emissivity-weighted
temperature ($T_X$).  Temperature is more directly related 
to a cluster's mass, but luminosity can also be mapped to
mass via an $L_X - T_X$ relation.  In either case, the cruical
link between models and observations is the mass-temperature
relation.

Recent analyses of high-redshift cluster temperature functions 
from the {\em Einstein} 
Extended Medium-Sensitivity Survey (EMSS) have shown that the
matter density of the universe probably lies in the range
$0.2 < \Omat < 0.7$ (Henry 1997; Donahue et al. 1998; Bahcall
\& Fan 1998; Eke et al. 1998; Donahue \& Voit 1999; but see
Blanchard \& Bartlett 1998; Viana \& Liddle 1999).  
Even though these conclusions are
based on rather few clusters, the systematic errors in these
measurements of $\Omat$ are comparable to the statistical
errors (Donahue \& Voit 1999).  With the flood of cluster
temperature measurements expected over the next few years from
{\em Chandra} and {\em XMM}, we will have the opportunity
to measure $\Omat$ much more precisely.  If we are to take full
advantage of these measurements, we will need to reduce the
systematic errors that currently exist in the modeling.
Here we concentrate on the uncertainties in the 
mass-temperature relation itself.

This paper analyzes the role of the mass-temperature relation
in characterizing cluster temperature evolution.  Section~2
outlines the formalism used to decribe the evolution of the
cluster mass function.  Section~3 investigates the physics
underlying the mass-temperature relation, showing that the
standard derivation of this relation from spherical top-hat
collapse is flawed and suggesting a new context for 
understanding the physical effects that govern this relation.
Section~4 discusses how to normalize the mass-temperature
relation and evaluates how severely uncertainties in this
normalization affect measurements of cosmological 
parameters.  Section~5 summarizes the
paper.

\section{Mass-Function Evolution}
\label{sec-theory}

Numerical simulations have shown that the Press-Schechter formalism
(Press \& Schechter 1974) and its various extensions (e.g., Lacey \&
Cole 1993) characterize gravitationally driven structure formation with surprising fidelity, particularly on cluster scales (e.g., Lacey 
\& Cole 1994; Eke, Cole, \& Frenk 1996; Bryan \& Norman 1998; 
Borgani et al. 1999a).  Because of its successes, this formalism 
is often used to relate observed cluster temperature and luminosity
functions to cosmological models (e.g., Henry \& Arnaud 1991; Eke \etal
1996; Borgani \etal 1999a).  While some recent large-scale
simulations have shown that the Press-Schechter formula might be
somewhat less successful at predicting the number density of the
most massive clusters (e.g., Governato et al. 1999),
we will assume in this paper that it is an exact description of the
cluster mass function, because here we are more concerned with 
analyzing systematic problems with the mass-temperature relation.

This section briefly outlines the formalism we will employ for
expressing the cluster mass function.  We derive expressions
for mass-function evolution in both open and flat universes,
and we assess the effects of a cosmological constant on
cluster evolution.  Many of these results have been derived
elsewhere; we compile them here as background for subsequent 
sections.

\subsection{Press-Schechter Formalism}

The Press-Schechter formalism for structure formation and 
its extensions describe how virialized objects grow from 
a field of initial density 
perturbations.  One defines $\delta({\bf x},t;M)$ to be the 
local fractional overdensity of the universe smoothed 
on mass scale $M$ and centered on comoving point ${\bf x}$ at 
time $t$. While these fluctuation amplitudes are linear, they 
grow in proportion to the function $D(t)$, which depends on 
$\Omat$ and $\Olam$, and their
rms amplitude on scale $M$ can be expressed as $\sigma(M) D(t)/
D(t_0)$.  Ultimately, some of these fluctuations grow non-linear,
and they are assumed to virialize when their amplitudes, extrapolated
from the linear regime according to $D(t)$, exceed some critical
threshold for virialization $\delta_c(t)$, which also depends on
$\Omat$ and $\Olam$.  One can then trace the merger history of 
a mass parcel beginning at location ${\bf x}$ from time 
$t_1$ to the present by keeping track of the largest $M$ for which
$\delta({\bf x},t_1;M)D(t)/D(t_1) > \delta_c(t)$.

Assuming that the perturbations are Gaussian, we can assess the number 
density of virialized objects with mass $>M$ by evaluating the quantity
$\nu_c(M,t) \equiv [\delta_c(t) / \sigma(M)]  [D(t_0)/D(t)]$, which is
the critical virialization threshold in units of the characteristic 
fluctuation amplitude.  The probability that a given mass parcel is part of a 
virialized structure of mass $>M$ is then equal to 
${\rm erfc}[\nu_c(M,t)/\sqrt{2}]$,
where ${\rm erfc}(q)$ is the complementary error function.  
Thus, the overall mass density in virialized objects exceeding 
mass $M$ is 
\begin{equation}
  \rho(>M) \; = \; \rho_0 \, {\rm erfc} \left( \frac {\nu_c} {\sqrt{2}} \right)
          \;  = \; \frac {2 \rho_0} {\sqrt{\pi}} 
                 \int_{\nu_c/\sqrt{2}}^{\infty} e^{-x^2} \, dx \; \; ,
\label{eq-intps}
\end{equation}
where $\rho_0$ is the mean mass density of the universe.
Differentiating this expression with respect to $M$ and dividing the
result by $M$ yields the familiar Press-Schechter formula for the 
comoving differential number density $dn$ of virialized objects 
within mass interval $dM$:
\begin{equation} 
  \frac {dn} {dM} (M,t) = \left( \frac {2} {\pi} \right)^{1/2} 
			\frac {\Omat \rho_{\rm cr,0}} 
		              {M^2}
			\left| \frac {d \ln \sigma} {d \ln M} \right| \,
			\nu_c(M,t) \, \exp[ -\nu_c^2(M,t)/2 ] \; \; ,
\label{eq-ps}
\end{equation}
where $\rho_{\rm cr,0} = 3 H_0^2 / 8 \pi G$ represents the present-day 
critical mass density.

The evolution of $dn/dM$ depends solely on $\nu_c(M,t)$.  
At present, we have $\nu_c(M,t_0) = \delta_c(t_0) / 
\sigma(M)$, and in principle, we can determine $\sigma(M)$ by fitting
equation~(\ref{eq-ps}) to the current distribution of cluster masses.
The function $\sigma(M)$ can be approximated by a power law with
index $\alpha = (n+3)/6$ on cluster scales, so that $\sigma(M) = 
\sigma_8 (M/M_8)^{-\alpha}$ with $M_8 = (H_0^2 \Omat /2G) (8 \, h^{-1} 
\, {\rm Mpc})^3 = 6.0 \times 10^{14} \, \Omat h^{-1} \, M_\odot$.
Holding $\sigma(M)$ fixed, we can project the current cluster mass
distribution backward in time as long as we know the functions 
$\delta_c(t)$ and $D(t)$.  More specific expressions
for $\sigma(M)$ describe how $n$ changes with the mass scale in
CDM-like models, but the analytical simplicity of the power-law
form better serves the illustrative purposes of this paper. 
 
\subsection{Cluster Evolution and $\Omat$}

Massive cluster evolution is very sensitive to $\Omat$ because
the number density of large clusters depends exponentially on
$\nu_c^2$ which is $\gg 1$.  Slight differences in the rate at 
which $\nu_c$ evolves therefore translate into large differences 
in cluster evolution (e.g., Oukbir \& Blanchard 1992; Eke \etal 
1996; Viana \& Liddle 1996).  
To illustrate how dramatic these differences can be, we will briefly 
outline the case of cluster evolution in an open universe with no 
cosmological constant.

Following Lacey \& Cole's (1993) treatment of perturbation growth
when $\Omat < 1$ and $\Olam = 0$ (see also the Appendix), we have
$\nu_c(M,t) = \omega(t) / \sigma(M)$ with
\begin{equation}
  \omega(t) \equiv \delta_c(t) D(t_0) / D(t) = \frac {3} {2} D(t_0)
             [1 + (t_\Omega/t)^{2/3}] \; \; ,
\end{equation}
where $t_\Omega = \pi \Omat / H_0 (1 - \Omat)^{3/2}$.
Normalizing the present value of $\nu_c$ at some fiducial mass scale
$M_0$, we can write
\begin{equation}
  \nu_c(M,t) = \nu_c(M_0,t_0) \left( \frac {M} {M_0} \right)^{\alpha} 
         \frac {1+(t_\Omega/t)^{2/3}} {1+(t_\Omega/t_0)^{2/3}} \; \; .
\end{equation}
Because the cluster number density at mass scale $M_0$ obeys
$dn/dM \propto \nu_c \exp (-\nu_c^2/2)$, we can use this equation
to gauge how rapidly clusters at this mass scale evolve.

\begin{figure}
\plotone{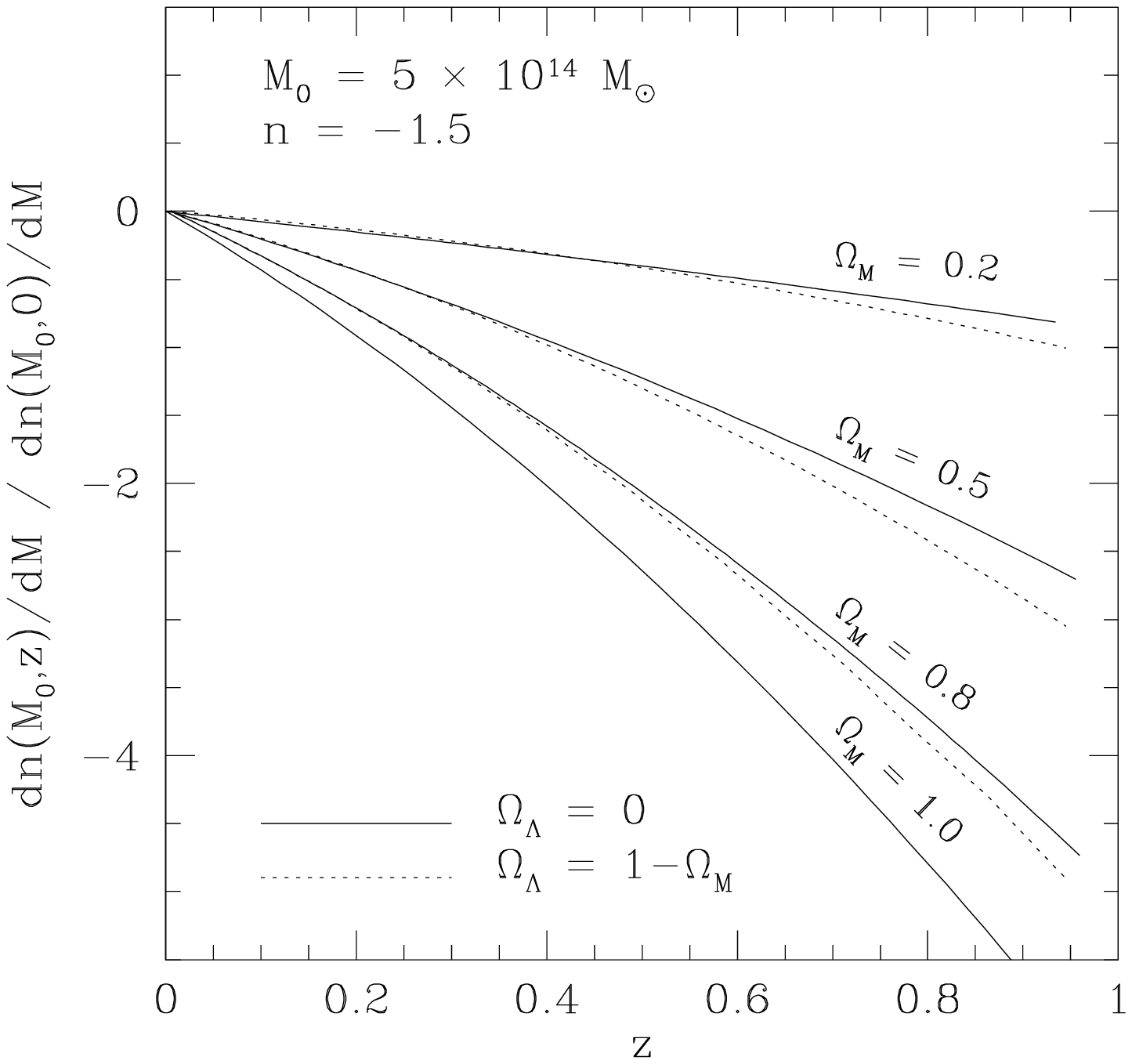}
\figcaption[mevxi.ps]{Expected evolution of massive clusters at mass 
scale $M_0$ for different values of $\Omat$.
\label{mevxi}}
\end{figure}

Figure~\ref{mevxi} illustrates how sensitively the number-density
evolution of massive clusters ($M_0 \approx 5 \times 10^{14} \, h^{-1} \,
M_\odot$) depends on $\Omat$.  For the purposes of this illustration, 
we have adopted the $\sigma_8$ fitting formulae of Eke \etal (1996) and 
have assumed that $n = -1.5$ on this mass scale.  In this example,
the number density of clusters in a universe with $\Omat = 0.8$
grows by about five orders of magnitude from $z = 1$ to the present,
a rapid rate of evolution that contrasts sharply with the single
order of magnitude expected in a universe with $\Omat = 0.2$. 
Changing the perturbation spectrum within observationally allowed
bounds changes the quantitative predictions somewhat, but the qualitative
conclusion remains the same: the number density of massive clusters
evolves much more rapidly for $\Omat \sim 1$ than for $\Omat \ll 1$.

\subsection{Cluster Evolution and $\Lambda$}

Clusters evolve slightly more rapidly in a flat, $\Olam >0$ universe 
than they do in an open, $\Olam = 0$ universe with the same value 
of $\Omat$ (e.g., Eke et al. 1996).  When a cosmological 
constant is operating, the universe's 
density remains close to critical later in time, promoting perturbation 
growth at lower redshifts.  However, cluster evolution is considerably
less sensitive to $\Olam$ than it is to $\Omat$.

In order to characterize cluster evolution in a flat universe with 
$\Olam > 0$ we require expressions for $D(t)$ and $\delta_c(t)$, 
which are derived in the Appendix.  From these expressions we
can construct the threshold function
\begin{equation}
  \omega(t) = - 9 \xi_c(t) D(t_0) \; \; ,
\end{equation}
where $\xi_c(t)$, defined in the the Appendix, is proportional 
to the specific energy of a 
perturbation that collapses at time $t$.  Normalizing $\nu_c$ as 
before at the mass scale $M_0$, we can write
\begin{equation}
  \nu_c(M,t) = \nu_c(M_0,t_0) \left( \frac {M} {M_0} \right)^{\alpha} 
         \frac {\xi_c(t)} {\xi_c(t_0)} \; \; .
\end{equation}
Plugging this expression into equation~(\ref{eq-ps}) then yields
the desired formulae for cluster evolution.

The dotted lines in Figure~\ref{mevxi} show how the number density
of clusters at mass scale $M_0$ evolves when $\Olam > 0$, for the
$\sigma_8$ normalization of Eke \etal (1996) and $n = -1.5$.  Note that the
rate of cluster evolution is quite insensitive to $\Olam$.
In this particular case, the best-fitting $\sigma_8$ for a flat 
universe is slightly higher than that for an open universe, which
almost compensates for the slightly more rapid rate of evolution
owing to $\Lambda$.  In general, the best-fitting $\Omat$ in an
open universe is $\sim 0.1$ higher than in a flat universe (e.g.,
Donahue \& Voit 1999).

\section{Theoretical Mass-Temperature Relations}

The Press-Schechter formalism conveniently describes the rate at which 
virialized objects of mass $M$ accumulate in the universe.  If we
could observe cluster masses directly, then comparisions between
Press-Schechter predictions and observed cluster evolution would
be simple.  Several types of observables, such as X-ray temperature,
cluster velocity dispersion, and weak lensing, are related
to cluster masses, but linking these quantities with the proper
Press-Schechter $M$ values requires careful attention.

This section focuses on the relation between cluster mass and
cluster temperature, the crucial relationship for linking X-ray
observations of clusters to models of structure formation. 
We first outline the observational evidence for a well-behaved 
mass-temperature relationship.  Then we analyze the standard
derivation of the mass-temperature relation, which is based on 
collapse of a spherical top-hat perturbation to an isothermal sphere.
This derivation yields a relation similar to the observed
relation, but it fails to conserve energy,
indicating that it omits important physical effects.  In an
effort to understand this relation more deeply, we present
a model for cluster formation, drawn from the merging-halo
formalism of Lacey \& Cole (1993), which accounts for the
fact that massive clusters accrete matter quasi-contiuously.
Analyzing clusters in this context enables us to identify
the physical effects that make up for the lack of energy
conservation.  The primary advantage of the continuous formation
model is that it more naturally reproduces the late-time evolution 
of clusters, and the section concludes by comparing 
predictions of cluster temperature evolution drawn from the 
continuous formation model with those from the spherical top-hat
model.

\subsection{Observational Evidence}

Simple scaling arguments suggest that the X-ray temperatures of 
clusters ($\tx$) should be directly related to their masses.  
One way to define a cluster's mass is to specify a characteristic
radius $r_\Delta$ within which the mean density is $\Delta$ times 
the critical density $\rhocr$, so that $M_\Delta = 4 \pi r_\Delta^3
\rhocr \Delta / 3$.  If all cluster potentials share the same density 
distribution, $\rho(r/r_\Delta)$, and the X-ray gas is isothermal, 
then $\tx \propto M_\Delta/r_\Delta \propto 
M^{2/3}_\Delta$.  Numerical simulations of cluster formation
demonstrate that this scaling ought to be remarkably tight,
with a scatter of only $15-20$\% (Evrard, Metzler, \& Navarro 1996;
Bryan \& Norman 1998).  These simulations also provide normalizations
for the mass-temperature relation that can be compared with 
actual clusters.

A recent observational investigation of the cluster mass-temperature 
relation at $z \lesssim 0.1$ by Horner, Mushotsky, \& Scharf (1999) 
supports the results of the simulations.  They show that cluster 
masses derived from velocity dispersions (Girardi \etal 1998) agree 
well with those inferred from ASCA temperatures (Fukazawa 1997), 
using the scaling law from Evrard \etal (1996) at $\Delta = 200$:
\begin{equation}
  M_{200} = (1.4 \times 10^{15} \, h^{-1} \, M_\odot)
                  \left( \frac {\tx} {10 \, {\rm keV}} \right)^{3/2}
                   \; \; .
\label{eq-mtobs}
\end{equation}
The scatter in the observed mass-temperature relation is $\sim
30$\%, but it decreases by a factor of 2 for the clusters with 
the highest numbers of measured galaxy redshifts, suggesting 
that the scatter intrinsic to the mass-temperature relation 
is probably quite small.  However, some concerns remain:
a handful of outliers deviate from the standard relation by
up to 50\% and the mass normalization one finds from hydrostatic
modeling of a subset of these clusters is 40\% smaller (Horner et al.
1999).

Similar comparisions at higher redshifts are more difficult, but the
available data indicate that the mass-temperature relation remains
well-behaved.  Hjorth, Oukbir, \& van Kampen (1998) have compared
masses derived from gravitational lensing analyses for 8 clusters 
at $0.17 \leq z \leq 0.54$ with the X-ray temperatures
of these clusters.  Their
best-fit mass-temperature relation agrees with the Evrard \etal
(1996) scaling law within the observational errors, and they conclude
that this scaling law can be used to measure masses to within 27\%.  
For clusters at even higher redshifts ($0.53 \leq z \leq 0.83$), 
Donahue \etal (1999) show that the observed relation between X-ray 
temperatures and cluster velocity dispersions remains consistent 
with the low-$z$ relation.

\subsection{Virial Mass and the Late-Formation Approximation}

The seemingly good behavior of the cluster mass-temperature 
relation is fortunate for those who wish to study cluster 
evolution with X-ray telescopes, but care must be taken when 
relating these temperature-derived masses to the virial masses 
demanded by the Press-Schechter formalism.  From the 
simulations and observations, we know that the 
mass within a specified density contrast is
straightforwardly related to temperature.  However, the density
contrast $\Delta_{\rm vir}$ corresponding to the virial radius 
depends in general on $\Omat$ and $\Olam$.  Thus, in order to 
characterize cluster evolution properly, we need to know how 
$\Delta_{\rm vir}(t;\Omat,\Olam)$ changes with time.

The usual approach to defining a cluster's virial mass is to 
approximate cluster formation with the evolution of a spherical 
top-hat perturbation (e.g., Peebles 1993).
Such a perturbation formally collapses to the origin at a 
particular moment ($t_c$) which is taken to be the moment 
of virialization.  The virialization time thus equals twice 
the time required for the perturbation to reach its turnaround 
radius ($r_{\rm ta}$).  A naive application of the virial theorem,
assuming that the perturbation is cold at maximum expansion, 
dictates that the cluster's final potential energy ought to be 
twice its potential energy at turnaround.  Hence, the cluster's
virial radius is assumed to be half its turnaround radius 
($r_{\rm vir} = r_{\rm ta} / 2$).
According to this prescription, $\Delta_{\rm vir}$ is a 
well-defined function of cosmic time and the parameters
$\Omat$ and $\Olam$ (Lacey \& Cole 1993; Kitayama \& Suto
1996; Oukbir \& Blanchard 1997).  In the case of $\Olam = 0$, 
this function can be concisely expressed as
$\Delta_{\rm vir} = 8 \pi^2 / (Ht)^2$,
where $H$ is the Hubble constant at time $t$.
If we additionally assume that each cluster we see at a given 
redshift $z$ has just reached the moment of virialization, an 
assumption known as the late-formation approximation, then 
$\mvir \propto \tx^{3/2} \rho_{\rm cr}^{-1/2} 
\Delta_{\rm vir}^{-1/2}$.

In a critical $\Omat = 1$ universe, the late-formation
approximation is valid because massive clusters develop rapidly 
at all redshifts; the effective moment of virialization is 
always close to the moment of observation.  However, 
in a universe with $\Omat < 1$, cluster formation is 
currently shutting down, and one must account for
differences between the moment of virialization and the moment
of observation.  This problem grows most severe at late
times in a $\Omat \ll 1$ universe, because the quantity
$\rhocr \Delta_{\rm vir}$ as determined via the late-formation 
approximation declines indefinitely.  The $\mvir$ associated 
with a given $\tx$ therefore rises steadily, even though cluster 
evolution has essentially stopped.  This spurious late-time 
evolution of the $\mvirtx$ relation is an undesirable
artifact of the late-formation approximation.

One approach to solving this problem is 
to account explicitly for the difference between the moment of 
virialization and the moment of observation in the context of 
a merging-halo formalism for cluster growth (Viana \& Liddle 1996;
Kitayama \& Suto 1996).  Another 
equivalent but mathematically simpler approach, which 
\S~\ref{contfrm} describes in detail, is to consider how 
the $\mvirtx$ relation should evolve in a population of 
clusters that gradually accrete their matter over an extended 
period of time, a more realistic scenario for the growth of 
very massive clusters (Voit \& Donahue 1998).

Calculating the normalization of the $\mvirtx$ relation under
the late-formation approximation is also somewhat problematic.  
Because a virialized cluster's potential is approximately 
isothermal, one would like to approximate it with a singular 
isothermal sphere, truncated at radius $r_{\rm vir}$, within 
which the mean density is $\rhocr \Delta_{\rm vir}$.  The 
one-dimensional velocity dipersion within such a potential 
is $\sigma_{\rm 1D}^2 = GM / 2r_{\rm vir}$ (Binney 
\& Tremaine 1987), which leads to the following relation between
virial mass and temperature:
\begin{eqnarray}
   k \tx & = & \frac  {G M^{2/3} \mu m_p} {2 \beta}
                      \left[ \frac {4 \pi} {3} 
                        \rhocr \Delta_{\rm vir} \right]^{1/3}
                     \nonumber \\
         & = &  (1.38 \, {\rm keV}) \beta^{-1} h^{2/3} M_{15}^{2/3}
                      \Delta_{\rm vir}^{1/3} 
                      \left[ \frac {\Omega_M} {\Omega_M(z)} \right]^{1/3}
                         (1+z)     \label{mtecf}
\end{eqnarray}
where $\beta = \mu m_p \sigma_{\rm 1D}^2 / k\tx$ and $\mu m_p
= 1 \times 10^{-24} {\rm g}$ is the 
mean mass per gas particle.

Comparisions of the $\mvirtx$ relation in equation~(\ref{mtecf}) 
with the masses and temperatures of simulated clusters indicate that 
$\beta^{-1} \approx 0.8-1$ (e.g., Bryan \& Norman 1998). This nearness of
$\beta$ to unity appears to validate the assumptions governing the
derivation of equation~(\ref{mtecf}), but the approximate
agreement between this equation and the simulations turns 
out to be something of a coincidence.  The total
energy of a collapsing spherical top-hat perturbation is
$-3GM^2/5r_{\rm ta}$.  After the collapsed perturbation
virializes into an isothermal sphere, a naive application of
the virial theorem that disregards boundary effects would place 
the total kinetic energy of the system at $3GM^2/5r_{\rm ta}$, 
corresponding to $\sigma_{\rm 1D}^2 = 2GM/5r_{\rm ta}$.  
The virial radius of the relaxed system would then be 
$5 r_{\rm ta} / 4$, a factor of 2.5 larger than assumed 
in the derivation of equation~(\ref{mtecf}), and its temperature
would correspondingly be 2.5 times lower.

In fact, truncation of a virialized system at some $r_{\rm vir}$ 
implies the existence of a confining pressure, unaccounted 
for in the top-hat collapse model, that alters the usual 
virial relationship between potential and kinetic energy 
(e.g., Carlberg, Yee, \& Ellingson 1997).  In the case of a 
singular isothermal sphere, the total kinetic energy is three 
times the absolute value of the total energy.  
Thus, energy-conserving collapse of a spherical top-hat 
perturbation into a pressure-truncated singular isothermal 
sphere should yield $\sigma_{\rm 1D}^2 = 6GM/5r_{\rm ta}$, 
implying $r_{\rm vir} = 5 r_{\rm ta} / 12$ (e.g., Shapiro, Iliev,
\& Raga 1999).  This result is close to the naive assumption 
that $r_{\rm vir} = r_{\rm ta} / 2$, but it is valid only 
if a confining pressure is applied at the virial radius.

\subsection{Continuously Forming Clusters}
\label{contfrm}

The inconsistencies in the top-hat, late-formation derivation 
of the $\mvirtx$ relation outlined above
indicate, perhaps unsurprisingly, that the top-hat collapse 
model excludes important physical effects that contribute 
to the normalization of the relation.  
In particular, the top-hat model accounts for neither the
energy and mass accumulation during the early stages 
of cluster formation nor the confining effects of matter
that continues to fall in, both of which significantly
increase the temperature associated with a given mass.
This section shows how these missing effects can be 
addressed in the context of a simple model in which 
massive clusters are allowed to form gradually, rather
than instantaneously.

In hierarchical models for structure formation, the growth
of the largest clusters is quasi-continuous.  The most massive
clusters are so rare that they almost never merge with another
cluster of similar size (e.g., Lacey \& Cole 1993).  Rather,
they grow by continually accumulating much smaller virialized
objects.  In the notation of \S~\ref{sec-theory}, their masses grow 
like $M \propto \omega^{-3/(n+3)}$ (Lacey \& Cole 1993; Voit
\& Donahue 1998).  Because each bit of infalling matter carries 
with it a specific energy $\epsilon$, we can compute the virial 
energy $-E$ of the cluster by integrating $E = - \int \epsilon dM$.  
The cluster temperature itself is proportional to $E/M$, so 
this integral also leads to a relation between virial
mass and temperature.

Voit \& Donahue (1998) treat the case of continuous cluster
growth when $\Omat < 1$ and $\Olam = 0$, finding that $M \propto 
x^{-3m/5}$, where $x = 1+(t_\Omega/t)^{2/3}$ and $m = 5/(n+3)$.  
Here we extend that calculation to
include the constant of proportionality between energy and mass.
Drawing on the Appendix, we express the specific energy of
infalling matter at time $t$ as
\begin{equation}
  \epsilon(t) = - \frac {1} {2} \left( \frac {2 \pi GM} 
       {t_\Omega} \right)^{2/3} (x-1) \; \; .
\end{equation}
Thus, we obtain
\begin{equation}
  \frac {E} {M} = 
       \frac {3} {10} \frac {m} {m-1} \left( \frac {2 \pi G} {t_\Omega}
       \right)^{2/3} M^{2/3} \left[ \left( \frac {t_\Omega} {t} \right)^{2/3}
        + \frac {1} {m} \right] \; \; .
\label{eq-vdnorm}
\end{equation}
In the limit of large $m$, which corresponds to the late-formation
approximation, this expression reduces to 
\begin{equation}
  \frac {E} {M} = - \frac {3} {5} \epsilon(t) \; \; ,
\end{equation}
which is identical to the $E/M$ ratio for a spherical top-hat
perturbation of mass $M$ that virializes at time $t$.
A similar procedure yields the mass-temperature relation in a
flat $\Olam > 0$ universe.  From the Appendix, we have $\epsilon(t) 
\propto M^{2/3} \xi_c(t)$ and $\omega(t) \propto  - \xi_c(t)$, 
giving
\begin{equation}
  \frac {E} {M} = - \frac {3} {5} \frac {m} {m-1} \epsilon(t) \; \; ,
\label{eq-vdlam}
\end{equation}
which again reduces to $ -3 \epsilon(t) / 5$ in the limit of large $m$.

Two factors in equation~(\ref{eq-vdnorm}) drive $E/M$ higher
than the late-formation value.  The $m/(m-1)$ factor, also
present in equation~(\ref{eq-vdlam}), accounts
for the effects of early infall; continuous cluster 
formation tends to create hotter clusters than top-hat
formation because more of the mass is assembled early, 
at a higher mean density.  For values of $n$ typical of 
cluster scales ($-2 \gtrsim n \gtrsim -1$), this factor ranges 
from 1.2 to 1.7.  The $1/m$ term in the bracketed factor
of equation~(\ref{eq-vdnorm})
accounts for the cessation of cluster formation when 
$t \gg t_\Omega$. At late times in an open universe, 
$E/M$ should remain constant, but in the late-formation 
approximation $E/M$ falls indefinitely because the fiducial 
density scale never stops dropping. 

Relating $E/M$ to temperature requires an expression for
the relationship between the total virial energy $-E$ and 
the total kinetic energy $E_K$.  When an external pressure
$P$ confines the boundary of a spherically symmetric virialized 
system, the appropriate form of the virial theorem can be written
\begin{equation}
  E_K = E + 4 \pi P r_{\rm vir}^3 \; \; . 
\end{equation}
If we take the velocity dispersion to be isothermal ($\sigma_{\rm 1D} = 
{\rm const.}$), then $P = \rho(r_{\rm vir}) \sigma_{\rm 1D}^2$, and
\begin{equation}
  E_K = \frac {\bar{\rho}} {\bar{\rho} - 2 \rho(r_{\rm vir})}  E 
         \; \; ,
\end{equation}
where $\bar{\rho}$ is the mean density within the virial radius.
In this formulation, the ratio $E_K/E$ depends on the shape
of the potential within $r_{\rm vir}$.  If the local density
is negligible at $r_{\rm vir}$, then the confining pressure 
is effectively zero and $E_K = E$.  If the potential 
strictly obeys $\rho \propto r^{-2}$, then $E_K = 3 E$.

Because we wish to derive an approximate mass-temperature 
relation for comparison with observations and simulations 
giving the temperature and mass of a cluster within
$r_{200}$, let us compute $E_K/E$ for the ``universal'' 
density profile of Navarro, Frenk, \& White (1997), truncated
at $r_{200}$, under the assumption that $\sigma_{\rm 1D}$ is 
constant.  The ratio of mean density to local density within
$r_{200}$ is then
\begin{equation}
  \frac {\bar{\rho}} {\rho(r_{200})} = 3 \frac {(1+c)^2} {c^2}
          \ln \left[(1+c) - \frac {c} {1+c} \right] \; \; ,
\end{equation}
where $c$ is a parameter that quantifies the concentration of
matter toward the cluster's center.  Simulations by Eke, Navarro, 
\& Frenk (1998) show that $c \approx 4 - 6.5$ for clusters 
at $0 < z < 1$ in a $\Omat = 0.3$, $\Olam = 0.7$ universe, 
and simulations of cluster formation by Navarro \etal (1997) 
show that the most massive clusters in critical universes
exhibit similar levels of concentration.  Values of 
$c \sim 5$ lead to density contrast factors $\approx 4$ 
and $E_K/E \approx 2$.

An intriguing alternative approach by Shapiro \etal (1999)
investigates the post-collapse structures of clusters by
seeking the minimum-energy solution among a family of
non-singular truncated isothermal spheres.  The minimum-energy
solution turns out to closely resemble the self-similar
spherical infall solution of Bertschinger (1985).  Because
the truncation radius of this minimum-energy isothermal
sphere is nearly equal to the accretion-shock radius of
the infall solution, Shapiro \etal (1999) suggest that
continual infall naturally maintains the confining pressure 
on the virialized isothermal sphere.  In this model, the 
density contrast factor at the truncation radius is 3.73, 
and $E_K/E = 2.17$.

Taken together, the effects of early accretion and pressure
confinement make up for the lack of energy conservation 
in the top-hat, late-formation derivation of equation~(\ref{mtecf}).
In the early-time limit ($t \ll t_\Omega$), equation~(\ref{eq-vdnorm})
yields the following mass-temperature relation:
\begin{equation}
   k \tx = \left[ \frac {2} {5} \frac {m} {m-1} \frac {E_K} {E} \right]
               \frac  {G M^{2/3} \mu m_p} {2 \beta}
                      \left( \frac {4 \pi} {3} 
                        \rhocr \Delta_{\rm vir} \right)^{1/3}
            \; \; .
\end{equation}
When $n \approx -2$ and $E_K/E \approx 2$, the prefactor in
brackets is close to unity, making this expression nearly
identical to equation~(\ref{mtecf}).

The lesson here is that the assumptions underlying 
equation~(\ref{mtecf}) are physically unsound.
The approximate agreement between the $\mvirtx$ normalization 
derived via the top-hat collapse model and those derived
from simulations and observations is largely coincidental.
As long as $\Omat$ is not very small, the time-dependent
factors in equations~(\ref{mtecf}) and (\ref{eq-vdnorm}) do not 
differ by a large factor.  However, because equation~(\ref{eq-vdnorm}) 
more faithfully reflects the behavior of cluster formation 
in all the appropriate limits, we prefer to base the $\mvirtx$ 
relation on the continuous-formation model.

\subsection{Late-Formation vs. Continuous-Formation}

The $\mvirtx$ relations in equation (\ref{mtecf}), derived 
using the late-formation approximation, and equation 
(\ref{eq-vdnorm}), derived using the 
continuous-formation approximation, differ
in both normalization and time-dependent behavior.
The following section will discuss the importance of properly 
normalizing the $\mvirtx$ relation.  Here we wish to examine how 
differences in time-dependence alone translate into different 
predictions for cluster evolution.  In order to isolate the 
time-dependent behavior, we can identically normalize both 
$\mvirtx$ relations to equation (\ref{eq-mtobs}) at $z=0$ 
and compare the resulting cluster temperature functions.

\begin{figure}
\plotone{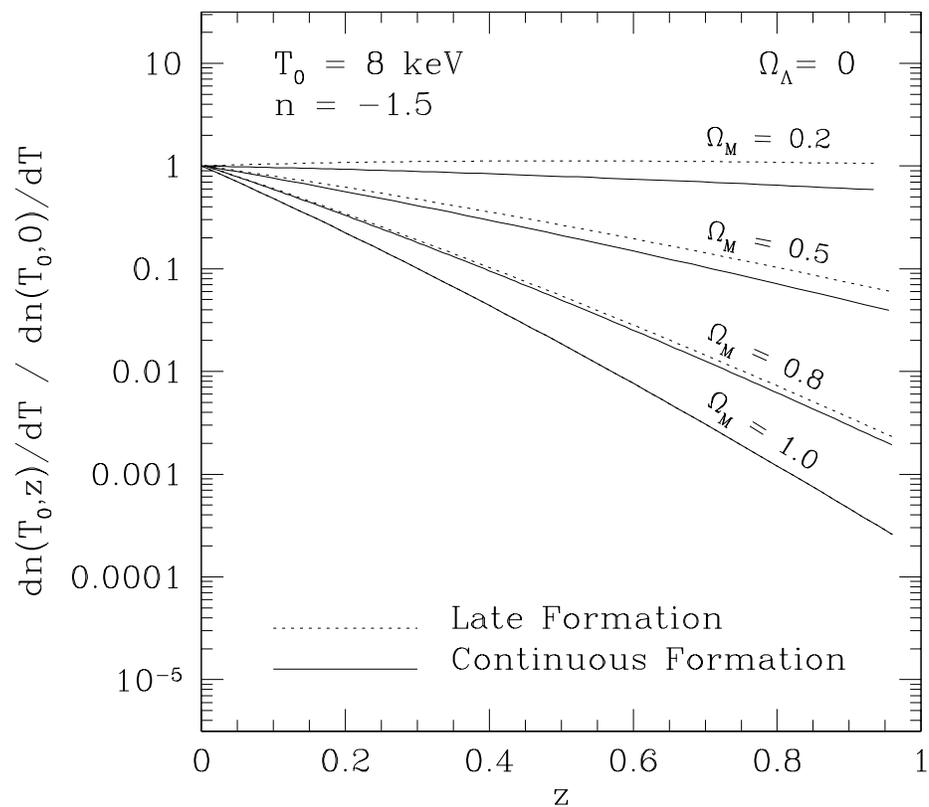}
\figcaption[tevcmp.ps]{Expected evolution in the comoving number density
of 8 keV clusters for the late-formation approximation and the
continuous formation approximation.  If $\Omat \approx 0.2$, the 
best-fitting $\Omat$ for the late-formation and continuous-formation
cases can differ by $\sim 0.1$.
\label{tevcmp}}
\end{figure}

Figure~\ref{tevcmp} shows the evolution of the temperature 
function for 8~keV clusters, given the $\sigma_8$ normalization 
of Eke \etal (1996).  Because the $\mvirtx$ relation evolves 
less strongly in the continuous-formation case, the rise in 
cluster temperature at a given mass as $z$ increases does 
not compensate as fully for the drop in the number of clusters 
at that mass.  The evolution of the temperature function at a 
given value of $\Omat$ is therefore stronger in the 
continuous-formation case (see also Viana \&
Liddle 1999).  Correspondingly, the best-fitting $\Omat$ to a
given observed amount of cluster evolution will be lower.
In this particular case, the difference amounts to $\sim 0.1$ 
in $\Omat$ for a best-fitting $\Omat \approx 0.3$.

Because the statistical errors in $\Omat$ derived from cluster
temperature evolution are also $\sim 0.1$, this discrepancy 
between the late-formation and continuous-formation approximations 
will need to be resolved if we are to take full
advantage of the cluster temperature measurements expected
from {\em Chandra} and {\em XMM}.  The best way to proceed
will be to test how well these temperature-function predictions
represent the results of large-scale structure simulations.  
However, cluster temperature functions will have to
be extracted directly from the simulated data, without resorting 
to a mass-temperature conversion step, presumably by using
the cluster-particle velocity dispersion as a surrogate
for cluster temperature.

\section{Normalization of the $\mvirtx$ Relation}

Both the top-hat and continuous-formation derivations of
the $\mvirtx$ relation given in the previous section 
have holes which must be plugged with knowledge gained from 
simulations.  In fact, any such spherically symmetric representation
glosses over aspects of cluster formation that are inherently 
three-dimensional.  Thus, it seems wise to normalize these 
analytical expressions to the results of numerical simulations.  
This procedure appears simple enough, but one must bear in mind 
that the normalization depends on $\Omat$ and that simulations 
have been done only for a few particular values of $\Omat$.  
Here we explain how we choose to normalize the $\mvirtx$ relation 
then investigate the consequences of an offset 
in the normalization.

\subsection{Normalizing to Simulations}

Because of the good agreement between the observational 
compilation of Horner \etal (1999) and the simulations of 
Evrard \etal (1996), we would like to normalize the
$\mvirtx$ relation accordingly.  Applying the time-dependence
factors derived for continuously forming clusters to the
empirical mass-temperature relation in equation (\ref{eq-mtobs})
thus gives
\begin{equation}
  k \tx = (8.0 \, {\rm keV})  \left( \frac {M} {10^{15} \, h^{-1} \, M_\odot}
         \right)^{2/3} 
       \left[ \frac {(t_\Omega/t)^{2/3} + 1/m} 
                    {(t_\Omega/t_0)^{2/3} + 1/m} \right] \; \; 
\end{equation}
and
\begin{equation}
  k \tx = (8.0 \, {\rm keV}) \left( \frac {M} {10^{15} \, h^{-1} \, M_\odot}
       \right)^{2/3} \frac {\xi_c(t)} {\xi_c(t_0)} \; \; .
\end{equation}
for open and flat universes, respectively.

\begin{figure}
\plotone{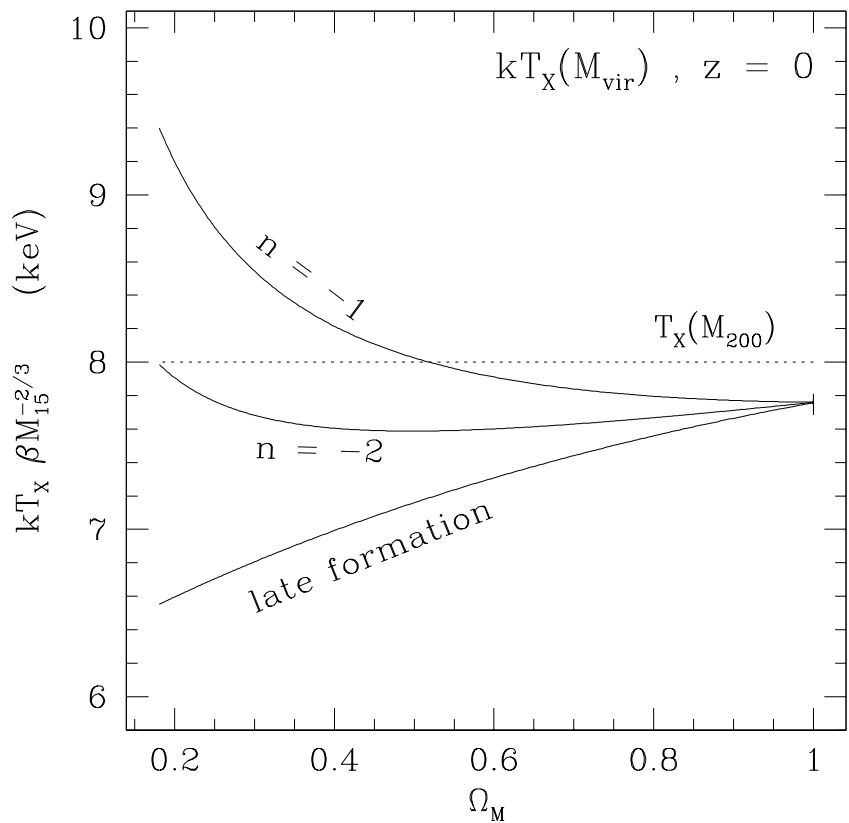}
\figcaption[mtnorm.ps]{Dependence of the $\mvirtx$ normalization
on $\Omat$.  The dotted line represents the temperature of a
cluster containing $10^{15} \, h^{-1} \, M_\odot$ cluster within
the radius $r_{200}$, as predicted by the simulations of Evrard
\etal (1996).  The solid line labeled ``late formation'' shows 
the temperatures predicted by $\mvirtx$ relation of Eke \etal (1996),
and the other two solid lines represent the $\mvirtx$ relations
from Voit \& Donahue (1998) for $n = -1$ and $-2$. 
\label{mtnorm}}
\end{figure}

Figure~\ref{mtnorm} compares this empirical normalization with 
the normalizations of the $\mvirtx$ relations derived by Eke et al.
(1996) and Voit \& Donahue (1998) at $z = 0$, assuming $\Olam = 0$.  
The dashed line indicates the normalization given in
equation~(\ref{eq-mtobs}), which is presumed to be independent 
of $\Omat$.  The curve labeled ``late formation''
shows the normalization of equation (\ref{mtecf}), derived from
the top-hat, late-formation model.
This normalization drops steadily with decreasing $\Omat$ because
the density contrast factor $\Delta_{\rm vir}$ grows smaller as
$\Omat$ declines.  Clusters modelled in this way are therefore
less compact and cooler than one would expect from the critical
density alone.  When $\Omat = 1$, this normalization is only 4\%
below the empirical value, but if $\Omat = 0.2$, it lies 20\% below
this value, corresponding to mass discrepancies of 6\% and 30\%,
respectively.

The behavior of the normalizations derived from continuous-formation
models is more complicated.  Voit \& Donahue (1998) normalized these
relations to the Eke et al. (1996) relation at $\Omat = 1$ to simplify
comparisons.  For $\Omat \lesssim 1$, they are lower than at 
$\Omat = 1$ for the same reason as in the late-formation model.
However, if $\Omat \ll 1$, cluster formation happened long before
$z \approx 0$, when the universe was considerably denser.  Clusters
are therefore denser and hotter than one would expect from the current 
critical density.  As a result, the temperature normalization of 
the $n = -2$ case deviates by less than 10\% over the range 
$0.2 < \Omat < 1.0$.  In the $n = -1$ case the normalization is 
actually 18\% higher than the empirical value (27\% in mass) at 
$\Omat = 0.2$.

This comparison illustrates why the procedure of normalizing 
the $\mvirtx$ relation to simulations is imperfect.  In general,
we expect this normalization to vary with $\Omat$ in a way that
depends on $n$.  Given that we have simulations for only a handful
of cosmological models, how do we unambiguously normalize these
relations?  Furthermore, different choices for extrapolating this 
normalization to other values of $\Omat$ can lead to normalizations
that differ by as much as 40\% in temperature (60\% in mass) at
a given value of $\Omat$.  Because of these uncertainties
in the normalization of the $\mvirtx$ relation, it is important
to understand how offsets in the normalization affect cosmological
parameters derived from the cluster temperature function.

\subsection{Normalization Offset and $\sigma_8$}

The $\mvirtx$ relation is invoked twice in the usual derivation
of $\sigma_8$ from the low-redshift temperature function.  In both
instances, an overestimate of the mass associated with a given 
temperature drives the best-fitting value of $\sigma_8$ higher.
For example, a 50\% offset in the mass normalization changes $\sigma_8$ 
by about 15\%.  Systematic uncertainties in the $\mvirtx$ relation
therefore lead to systematic uncertainties in $\sigma_8$, limiting
the usefulness of the temperature function as a tool to measure
the perturbation amplitude.

The first place the $\mvirtx$ relation enters is in the conversion
of the theoretical cluster mass function $dn/dM$ in equation (\ref{eq-ps})
to the cluster temperature function
\begin{equation} 
  \frac {dn} {dT} (T,t) = \frac {3} {2} \left( \frac {2} {\pi} \right)^{1/2} 
			\frac {\Omat \rho_{\rm cr,0}} 
		              {T \, M(T,t)}
			\left| \frac {d \ln \sigma} {d \ln M} \right| \,
	     \nu_c[M(T,t),t] \, \exp \{ -\nu_c^2[M(T,t),t)]/2 \} \; \; .
\label{eq-dndt}
\end{equation}
An $\mvirtx$ relation that overestimates $M(T)$ by a fractional amount
$\delta_M$ will underpredict the density $dn/dT$ by the same fractional
amount.  This source of error drives the best-fitting value of $\nu_c$
lower by a fractional amount $\delta_\nu \approx - \delta_M / (\nu_c^2 - 1)$.

Once $\nu_c(T)$ has been derived over a given range of temperatures,
one can determine $\sigma(T) = \delta_c(t_0) / \nu_c(T)$.  These $\sigma$ 
values will be too high by a fraction $\approx \delta_M / (\nu_c^2 - 1)$
if there is a normalization offset.
Conversion of $\sigma(T)$ to $\sigma(M)$ contributes another term to
the systematic error budget.  If $M(T)$ is overestimated, the mismapping
of temperature to mass inflates $\sigma_8$ by a fractional amount
$\approx \alpha \delta_M$.

As an example of these effects, consider the derivation of $\sigma_8$
from the abundance of $> 5$~keV clusters.  Markevitch (1998) finds that 
the number density of such clusters at $z \sim 0$ is $\approx 7.0 \times 
10^{-7} \, h^3 \, {\rm Mpc}^{-3}$.  According to the mass-temperature 
relation in equation (\ref{eq-mtobs}), the mass of a 5~keV cluster is 
$5.0 \times 10^{14} \, h^{-1} \, M_\odot$, leading to an overall mass 
density $\rho(> 5 \, {\rm keV}) \approx 2.3 \times 10^{-32} \, h^2 
\, {\rm g \, cm^{-3}}$ in such objects.  Plugging this value into equation 
(\ref{eq-intps}) and solving for $\nu_c$ yields $\nu_c(5 \, {\rm keV})
\approx 3.2$ for $\Omat = 1$ and $\nu_c(5 \, {\rm keV}) \approx 2.8$ 
for $\Omat = 0.3$, numbers that are consistent with more rigorous fits
to cluster temperature data using a similar mass-temperature relation
(Donahue \& Voit 1999). 

Conversion of these $\nu_c$ values to $\sigma_8$ values depends on the
shape of the initial perturbation spectrum and the value of the 
virialization threshold $\delta_c$.  For the purposes of this analysis,
we will assume that $\delta_c$ is known perfectly, implying that
$\sigma(5.0 \times 10^{14} \, h^{-1} \, M_\odot) \approx 0.53$ for 
$\Omat = 1.0$ and $\sigma(5.0 \times 10^{14} \, h^{-1} \, M_\odot) 
\approx 0.59$ for $\Omat = 0.3$.  Extrapolating along the mass
spectrum assuming, for example, that $n = -1.5$ then leads to $\sigma_8
\approx 0.51$ for $\Omat = 1.0$ and $\sigma_8 \approx 0.76$ for
$\Omat = 0.3$.

Now let us inflate the normalization of the mass-temperature relation 
by 50\% in mass.  The overall mass density in objects $> 5$~keV rises 
to $\rho(> 5 \, {\rm keV}) \approx 3.5 \times 10^{-32} \, h^2 
\, {\rm g \, cm^{-3}}$, yielding $\nu_c(5 \, {\rm keV}) \approx 
3.1$ for $\Omat = 1$ and $\nu_c(5 \, {\rm keV}) \approx 2.7$ for 
$\Omat = 0.3$.  Because 5~keV corresponds to $7.5 \times 10^{14}
\, h^{-1} \, M_\odot$ under the alternative normalization, we 
now have $\sigma(7.5 \times 10^{14} \, h^{-1} \, M_\odot) \approx 
0.54$ for $\Omat = 1.0$ and $\sigma(7.5 \times 10^{14} \, h^{-1} 
\, M_\odot) \approx 0.61$ for $\Omat = 0.3$.  Extrapolation to
$\sigma_8$, again assuming $n = -1.5$, gives $\sigma_8 \approx
0.58$ for $\Omat = 1.0$ and $\sigma_8 \approx 0.87$ for $\Omat
= 0.3$. Note that these systematic changes, of order 15\%, exceed 
the typically quoted measurement errors for $\sigma_8$ at a fixed 
value of $\Omat$.  

Additional uncertainty in the derived value of $\sigma_8$ can
arise from uncertainty in the slope $n$ of the perturbation spectrum.
For example, the $n \approx -2.3$ slope derived from the Markevitch
cluster sample (Markevitch 1998; Donahue \& Voit 1999) leads to a 
considerably lower derived value of $\sigma_8$ when $\Omat \ll 1$.
In the case of $\Omat = 0.3$, a measurement of $\sigma(5.0 \times 
10^{14} \, h^{-1} \, M_\odot) \approx 0.59$ extrapolates to 
$\sigma_8 \approx 0.66$, corresponding to clusters of temperature
$\sim 2.5 \, {\rm keV}$, below the temperature limit of the sample.
Blanchard et al. (1999) have recently argued that the Markevitch
sample is incomplete at low temperatures, implying $n > -2.3$.  That 
is probably why the best-fitting $\sigma_8$ values of Donahue \& Voit 
(1999), based primarily on the Markevitch clusters, seem unusually 
low. (Errors on $\sigma_8$ quoted in that paper refer only to the 
statistical errors in $\sigma_8$ at the best-fitting value of $\Omat$.)
Ideally, one would like to measure $\sigma_8$ from the number 
density of clusters at temperatures corresponding to the appropriate 
mass scale, but to do this, one first needs a reliable $\mvirtx$ 
relation, in addition to a well constrained value of $\Omat$.

The upshot of this analysis is that $\sigma_8$ values derived 
from the cluster temperature function contain systematic errors
that depend on the mass-temperature relation.  These systematic
errors are currently comparable to the measurement errors.  
Until the $\mvirtx$ relation is better understood, $\sigma_8$
values derived from the cluster temperature function will have
to be treated with caution.  Conversely, predictions of cluster
temperature functions that invoke $\sigma_8$ values derived from
other kinds of data will also contain systematic errors owing 
to normalization uncertainties in the $\mvirtx$ relation.

\subsection{Normalization Offset and $\Omat$}

The systematic problems discussed above in relating $\sigma_8$ 
to $dn/dT$ need not lead to unwarranted pessimism about deriving 
$\Omat$ from the evolution of $dn/dT$.  The key quantity in 
establishing the rate of cluster temperature evolution is not
$\sigma_8$, but rather $\nu_c(T,t_0)$, whose systematic errors
are considerably smaller, $\sim 5$\% instead of $\sim 15$\%.  
In order to evaluate the impact of a $\mvirtx$ normalization 
offset on predictions of temperature evolution, we will first 
analyze the case of $\Omat = 1$, then consider how a normalization
offset affects measurements of $\Omat$.

When $\Omat = 1$, the time-dependent parts of $M(T,t)$ and 
$\nu(T,t)$ simplify to $M \propto (1+z)^{-3/2}$
and $\nu_c \propto (1+z)^{(2-3\alpha)/2}$.  The amount
of cluster evolution at a fixed temperature $T$ can
therefore be written as
\begin{equation}
  C(T,z) = \frac {\frac {dn} {dT} (T,z)} {\frac {dn} {dT} (T,0)}
         = (1+z)^{(5-3\alpha)/2} 
              \exp \{ - \frac {\nu_c^2(T,0)} {2} [(1+z)^{(2-3\alpha)} -1]
                   \} \; \; .
\end{equation}
A fractional overestimate $\delta_M$ of cluster masses thus leads
to an underestimate of the amount of evolution by a factor 
$\sim \exp \{ \delta_M [(1+z)^{2-3\alpha} - 1 ] \}$.
If the overestimate of cluster masses is 50\%, this factor
amounts to a 20\% underestimate of evolution at $z = 0.3$ and 
a 70\% underestimate at $z = 0.8$ for $n = -1.5$.  

These uncertainties are relatively minor compared to the expected
amount of evolution.  For example, if $n = -1.5$ and $\nu_c(T,z=0) 
\approx 3.2$, we expect $C(T,z=0.3) \approx 0.3$ and $C(T,z=0.8) 
\approx 0.03$.  Larger values of $\nu_c$, characteristic of hotter
clusters, lead to even more evolution.  Because these evolution
predictions are over an order of magnitude larger than the
systematic errors, $\mvirtx$ normalization discrepancies do
not seriously affect the conclusion that cluster temperature
evolution rules out $\Omat = 1$, particularly when clusters
at $z > 0.5$ are included. 

\begin{figure}
\plotone{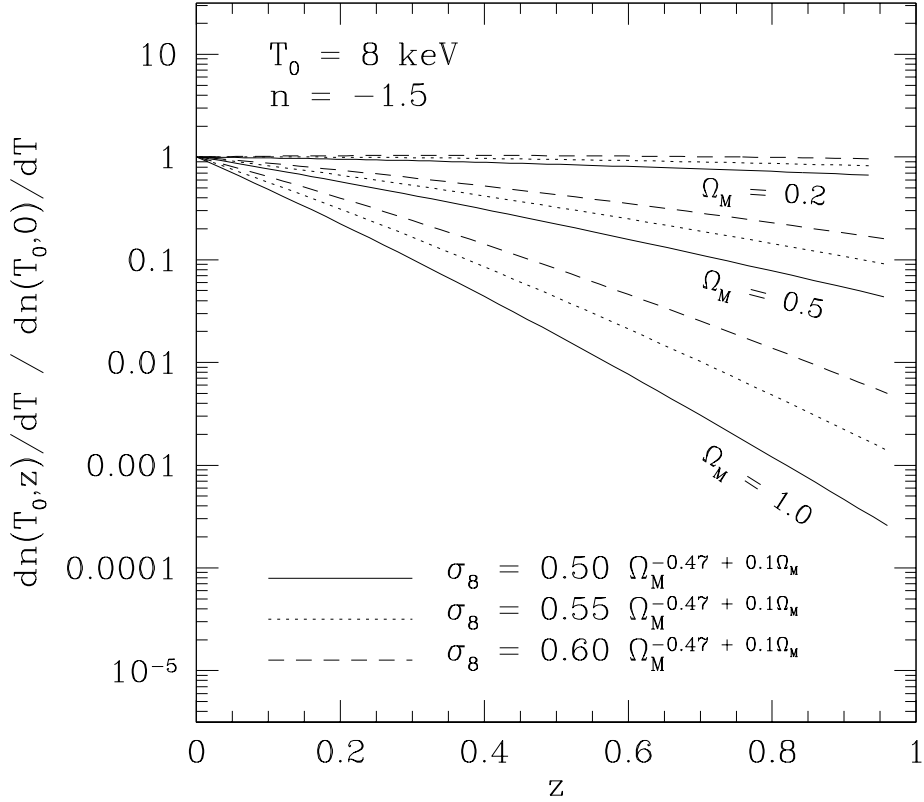}
\figcaption[tevxisig.ps]{Expected evolution in the comoving number density
of 8 keV clusters for $0.5 \leq \sigma_8 \Omat^{0.47 - 0.1\Omat} \leq 0.6$.
Because of uncertainties in the $\mvirtx$ relation, $\sigma_8$ is not tightly 
contrained by the cluster temperature function.  If these uncertainties are
taken at face value, they result in a considerable spread in predictions
for cluster evolution.
\label{tevxisig}}
\end{figure}

\begin{figure}
\plotone{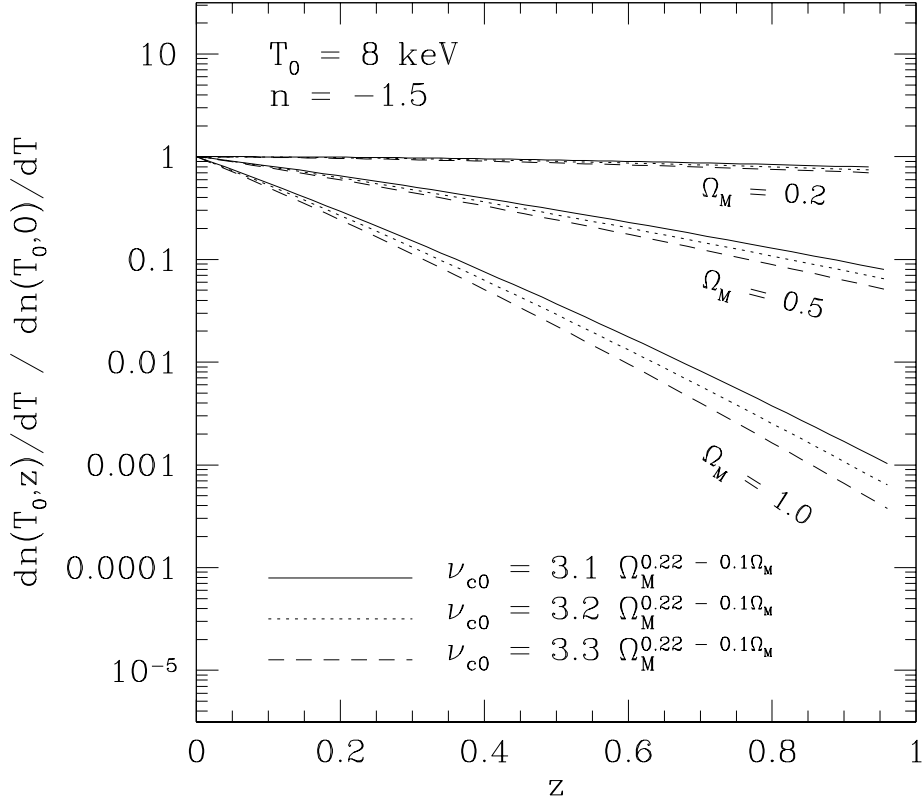}
\figcaption[tevxinu.ps]{Expected evolution in the comoving number density
of 8 keV clusters over the range of $\nu_c$ values corresponding to the 
observed low-redshift cluster temperature function at 5~keV.  Despite
the uncertainties in the $\mvirtx$ relation, the parameter 
$\nu_{c0} = \nu_c( 5 \, {\rm keV}, z = 0)$ is fairly well constrained.  A 50\%
change in the mass-normalization, or correspondingly, a 50\% change
in the comoving number density of 5~keV clusters, changes the best
fitting $\nu_c$ by $\sim 5$\%.  The resulting spread in 
predictions for cluster evolution is not as severe as one might have
guessed from the uncertainties in $\sigma_8$. 
\label{tevxinu}}
\end{figure}

Partially because of the potentially significant uncertainty in $\sigma_8$,
certain authors have been cautious about the conclusions that can
be drawn about $\Omat$ from cluster temperature evolution
(Colafranceso, Mazzotta, \& Vittorio 1997; Viana \& Liddle 1999;
Borgani \etal 1999b).  However, maximum likelihood methods of 
determining $\Omat$ that compare the cluster temperature function 
at $z \approx 0$ directly with the cluster temperature function 
at higher redshifts (e.g., Henry 1997; Donahue \& Voit 1999) 
can obtain stronger constraints on $\Omat$ because they are 
differential measurements in which much of the uncertainty
in $\sigma_8$ and $\delta_c$ cancels.
Figure~\ref{tevxisig} shows the cluster evolution predictions
that result when a representative range of $\sigma_8$ values is considered.
Here we allow $0.5 \leq \sigma_8 \Omat^{0.47 - 0.1 \Omat} \leq 0.6$.
At the $z \approx 0.3$ redshift of the Henry (1997) clusters
the prediction of the high-$\sigma_8$, $\Omat = 1$ model is only 
a factor of two lower than the low-$\sigma_8$, $\Omat = 0.5$ model, 
underscoring the need to identify systematic sources of uncertainty
in $\sigma_8$ before deriving evolutionary predictions for clusters
from it. 
However,
Figure~\ref{tevxinu} paints a somewhat rosier picture.  Here we
allow $\nu_{c0} = \nu_c(5 \, {\rm keV}, z=0)$ to span a range that
corresponds to a factor of two range in the mass normalization
at 5~keV, or equivalently, a factor of two range in the number
density of 5~keV clusters at $z = 0$.  The resulting systematic
uncertainty in the best-fitting $\Omat$ is $\lesssim 0.1$.

\subsection{Toward Greater Precision}

When using Press-Schechter methods to model cluster evolution,
one should always keep in mind that they are useful because
they efficiently approximate numerical simulations.
Our confidence in these methods is rooted in the fact that
they reproduce the mass function of simulated clusters reasonably
well.  Less work has been done on comparisons of 
simulated cluster temperature functions with 
temperature functions derived from Press-Schechter mass
functions using an $\mvirtx$ relation (e.g., Bryan \& Norman
1998, Pen 1998).  

Given the ambiguities surrounding the $\mvirtx$ 
relation and the very definition of a cluster's mass, 
the most robust way to model the evolution of $dn/dT$ 
would seem to be with a version of the Press-Schechter 
formalism that describes the temperature function directly.  
In such a scheme, one would separate the function $\nu_c(T,t)$
that determines the evolution of $dn/dT$ into a temperature-dependent
part $\nu_c(T,0)$ and a time-dependent part $g(t;\Omat,\Olam)$. 
In principle, $g(t;\Omat,\Olam)$ could be derived from a grid 
of simulated cluster temperature functions.  However, creating 
such a simulation set would be very expensive.  The simulation
volume would need to be extremely large to obtain adequate 
statistics on rare, high-temperature clusters.

Maximum likelihood fits to cluster temperature surveys are essentially
seeking the best-fitting $\nu_c(T,z) = \nu_c(T,0) g(z; \Omat, \Olam)$.
Figure~\ref{tevxinu} shows that this technique 
is fairly robust with respect to systematic uncertainties in the 
$\mvirtx$ normalization.  Statistical uncertainties in the normalization 
and slope of $\nu_c(T,0)$ are handled naturally by the maximum likelihood 
method.  Insofar as the dependence of $\nu_c(T,z)$ on cosmological 
parameters is accurate, this technique currently has the potential 
to deliver values of $\Omat$ that are accurate to $\lesssim 0.1$.  
However, it remains to be seen how accurately our assumed forms for 
$g(z; \Omat, \Olam)$ reproduce the results of large-scale
clustering simulations.

\section{Summary}

X-ray surveys of distant clusters are placing increasingly more
stringent constraints on $\Omat$.  The lack of extreme evolution
in the cluster temperature function strongly indicates that 
$\Omat < 1$.  One of the crucial ingredients in placing such
constraints on $\Omat$ is the $\mvirtx$ relation that converts 
cluster temperatures to cluster masses, enabling us to relate 
X-ray temperature surveys to theoretical models for cluster 
formation.  If we are to extract accurate values of $\Omat$
from the larger cluster temperature surveys expected from
{\em Chandra} and {\em ASCA}, we need to ensure that our
$\mvirtx$ relation faithfully describes cluster evolution when
coupled with Press-Schechter analysis.  To that end, this
paper has analyzed our current understanding of the cluster 
mass-temperature relation in an effort to identify the systematic 
errors it introduces into measurements of cosmological parameters.  

We find that the usual derivation of the $\mvirtx$ 
relation, which assumes that clusters form by spherical top-hat 
collapse and that we are observing them immediately after they 
formed, is physically inconsistent.  The rough agreement between 
the $\mvirtx$ normalization derived in this way and the 
normalization determined from numerical models of clusters is 
therefore somewhat coincidental.  
To obtain the proper normalization, one needs
to account both for the fact that much of a cluster's mass accreted
well before the moment we are observing it and for the non-zero 
density at $r_{200}$, which requires a surface
pressure term to be included in the virial theorem.

Because of these shortcomings of the spherical top-hat picture,
we advocate a more realistic scenario for deriving the $\mvirtx$
relation in which clusters form quasi-contiuously.  An expression 
for the $\mvirtx$ relation can be derived in the context of 
hierarchical merging, but the normalization of this relation 
depends on the concentration parameter $c$ of the cluster, 
which must be obtained from simulations.  The primary advantage
of this form for the $\mvirtx$ relation is that, unlike the
spherical top-hat model, it properly reproduces the cessation
of cluster evolution at late times if $\Omat < 1$.

Given the systematic uncertainties in setting 
the normalization of the $\mvirtx$ relation, we have investigated
their impact on the derivation of cosmological parameters
from the cluster temperature function.  Because two applications
of the $\mvirtx$ relation are needed to extract $\sigma_8$,
this parameter is particularly susceptible to uncertainties
in the mass-temperature normalization: a mass-normalization uncertainty
of 50\% leads to a 15\% uncertainty in $\sigma_8$. However,
only a single application of the $\mvirtx$ relation is needed
to extract $\Omat$, making it less vulnerable to normalization
uncertainties.  The systematic error in $\Omat$ owing to
uncertainties in the $\mvirtx$ relation is $\sim 0.1$.

Improvements in our understanding of the $\mvirtx$ relation
await comparisions of theoretically-derived cluster temperature
functions with structure-formation simulations large enough
to contain many hot clusters.  In essence, the $dn/dT$ expression
derived from the $\mvirtx$ relation and the Press-Schechter mass 
function is no more than an elaborate fitting formula for representing
the results of simulations.  The ideal mass-temperature relation
will therefore be the one that reproduces simulated temperature
functions or velocity-dispersion functions most accurately.
Even better would be a fitting formula that gives $dn/dT$ directly
without passing through murky intermediate steps involving ill-defined
cluster masses.  Until we have resolved these systematic uncertainties
in deriving $dn/dT$, our constraints on $\Omat$ from cluster
temperatures will not grow appreciably tighter.

\acknowledgements 

Megan Donahue's support, encouragement, and advice has been invaluable
to the author, who would also like to acknowledge Pat Henry,
Stefano Borgani, and the referee for helpful comments and NASA grants
NAG5-3257 and NAG5-3208 for partial support.

\begin{appendix}

\begin{center}
APPENDIX \\
CONTINUOUS CLUSTER GROWTH IN $\Omat < 1$ UNIVERSES
\end{center}

In the spirit of Gunn \& Gott (1972), we can idealize continuous 
cluster growth as occurring through the
sequential collapse and virialization of an infinite series of
concentric shells, each obeying the equation of motion
\begin{equation}
  \ddot{R} = - \frac {GM} {R^2} + \frac {\Lambda R} {3} \; \; ,
\end{equation}
where $R$ is the radius of the shell encompassing mass $M$, and $\Lambda$
is the cosmological constant.  The specific energy of the matter in the
shell is then
\begin{equation}
  \epsilon = \frac {\dot{R}^2} {2} - \frac {GM} {R} 
              - \frac {\Lambda R^2} {6} \; \; ,
\end{equation}
which remains constant until the shell virializes.  If the shell ever
reaches the critical radius $R_{\rm cr} = (3GM/\Lambda)^{1/3}$, cosmic 
repulsion dominates gravity from then on, and the shell never collapses.

We can cast the equation of motion for the shell into 
dimensionless form by defining
$x = R/R_{\rm cr}$, $\theta = \Lambda^{1/2} t$, and $\xi = \epsilon 
R_{\rm cr}^{-2} \Lambda^{-1}$, so that
\begin{equation}
  \frac {dx} {d\theta} = \left[ \frac {2} {3x} + 2 \xi 
                   + \frac {x^2} {3} \right]^{1/2}  \; \; .
\end{equation}
A particular shell will collapse if $x_0^3 + 6 \xi x_0 + 2 = 0$ for
some $x_0$ in the range $0 \leq x_0 < 1$.  Solving this cubic equation,
we find
\begin{equation}
  x_0 = 2^{3/2} |\xi|^{1/2} \cos \left( \frac {\alpha_\xi} {3} - 
              \frac {2 \pi} {3} \right) \; \; ,
\end{equation}
where $\alpha_\xi$ is defined by $\cos \alpha_\xi = - (8 |\xi|^3)^{-1/2}$
with $\pi/2 \leq \alpha_\xi \leq \pi$.
The shell therefore reaches its maximum radius $R_{\rm cr} x_0(\xi)$ at
a time $\Lambda^{-1/2} \theta_0(\xi)$, where
\begin{equation}
  \theta_0(\xi) = \int_0^{x_0(\xi)}  \frac {x^{1/2} \: dx}
         { \left( \frac {x^3} {3} + 2 \xi x + \frac {2} {3} \right)^{1/2} }
       \; \; .
\end{equation}
Because of the symmetry of the motion, the shell collapses to the origin
at $t_c(\xi) = 2 \Lambda^{-1/2} \theta_0(\xi)$, which we will take to be 
the time of virialization.

The overall scale factor of the universe obeys a similar equation of
motion, for which the specific energy of the background matter is
\begin{eqnarray}
  \epsilon_b & = & \frac {\dot{a}^2} {2} - \frac {H_0^2} {2} \Omat a^{-1}
                                         - \frac {H_0^2} {2} \Olam a^{2} \\
             & = & \frac {H_0^2} {2} (1 - \Omat - \Olam) \; \; .
\end{eqnarray}
The universe changes from decelerating to accelerating when $a = a_{\rm cr}
\equiv (\Omat / 2 \Olam)^{1/3}$, and with the definition $w = a / a_{\rm cr}$, the dimensionless equation of motion for the background expansion becomes
\begin{equation}
  \frac {dw} {d\theta} = \left[ \frac {2} {3w} + 2 \xi_b 
                   + \frac {w^2} {3} \right]^{1/2}  \; \; ,
\end{equation}
where $\xi_b = \epsilon_b a_{\rm cr}^{-2} \Lambda^{-1}$.

Taking advantage of the formal similarities between these equations of
motion, we can derive an expression for perturbation growth in an open
universe.  In the linear regime, $\delta \rho / \rho = -3 \delta w / w$,
where $\delta w = x - w$, and at any moment in time we have
\begin{equation}
  \int_0^x \frac {y^{1/2} \: dy} {\left( y^3 + 6 \xi y + 2 \right)^{1/2}}
 = \int_0^w \frac {y^{1/2} \: dy} {\left( y^3 + 6 \xi_b y + 2 \right)^{1/2}}
 \; \; ,
\end{equation}
At early times, when $x^3 + 2 \gg |6 \xi x|$ and $w^3 + 2 \gg 
|6 \xi_b w|$, we obtain
\begin{equation}
  \frac {\delta \rho} {\rho} = 9 (\xi_b - \xi) 
            \frac {(w^3 +2)^{1/2}} {w^{3/2}}
            \int_0^w \frac {y^{3/2} \: dy} {(y^3 + 2)^{3/2}} \; \; ,
\end{equation} 
and at the earliest times ($w \ll 1$), this expression reduces to
\begin{equation}
 \frac {\delta \rho} {\rho} = \frac {9} {5} (\xi_b - \xi) w \; \; .
\end{equation}
Perturbations at early times grow like $1/(1+z)$, as expected, and 
their amplitudes are proportional to the specific energy difference 
between the perturbation and the background.  Note that if the universe 
is flat, the background specific energy $\xi_b$ vanishes, and the 
perturbation amplitude within a shell is directly proportional to 
the shell's specific energy.

In the case of a vanishing cosmological constant, we can relate the 
perturbation amplitudes explicitly to their collapse times $t_c$.
When the cosmological constant is very small, we have $|\xi| \gg 1$
and $t_c \approx (\pi / 3\sqrt{2})|\xi|^{-3/2} \Lambda^{-1/2}$, so
a shell that collapses and virializes at time $t_c$ carries with
it a specific energy
\begin{equation}
  \epsilon = - \frac {1} {2} \left( \frac {2 \pi GM} {t_c} \right)^{2/3}
    \; \; .
\end{equation}
If we define $t_\Omega = \pi \Omat / H_0 (1 - \Omat - \Olam)^{3/2}$,
then at very early times,
\begin{equation}
  \frac {\delta \rho} {\rho} \approx \frac {3} {2} 
          \frac {(12 \pi)^{2/3}} {10} 
          \left( \frac {t} {t_\Omega} \right)^{2/3}
          \left[ 1 + (t_\Omega/t_c)^{2/3}  \right] \; \; .
\end{equation}
According to Lacey \& Cole (1993), the growth rate for linear perturbations 
in this limit is $D(t) \approx [(12 \pi)^{2/3}/10] (t/t_\Omega)^{2/3}$, 
and we retrieve their expression for $\delta_c(t)$:
\begin{equation}
  \delta_c(t) = \frac {3} {2} D(t) 
        \left[ 1 + (t_\Omega / t)^{2/3} \right] \; \; .
\end{equation}
Time therefore enters the Press-Schechter formula via the parameter 
$\nu_c \propto \delta_c(t)/D(t) \propto [1 + (t_\Omega / t)^{2/3}]$.

In the case of a flat universe with $\Omat < 1$, we have
\begin{equation}
  \frac {\delta \rho} {\rho} = - 9 \xi D(t)
\end{equation}
where
\begin{equation}
  D(t) = D[w(t)] =  \frac {(w^3 +2)^{1/2}} {w^{3/2}}
            \int_0^w \frac {y^{3/2} \: dy} {(y^3 + 2)^{3/2}} \; \; ,
\end{equation} 
in agreement with Eke et al. (1996).
Inverting the function $t_c(\xi)$ yields the function $\xi_c(t)$ giving
the specific energy $\epsilon(t) = R_{cr}^2 \Lambda \xi_c(t)$ of a shell 
that collapses to the origin at time $t$.  The collapse threshold
becomes
\begin{equation}
  \delta_c(t) = - 9 \xi_c(t) D(t) \; \; ,
\end{equation}
and time enters the Press-Schechter formula via $\nu_c \propto \delta_c(t)
/ D(t) \propto \xi_c(t)$.

\end{appendix}


\newpage

\end{document}